\documentclass[12pt,english,aps,nofootinbib,prb,preprint,superscriptaddress,showpacs,showkeys,12pt,sort&compress,longbibliography]{revtex4-1}
\usepackage{ae,aecompl}
\usepackage[T1]{fontenc}
\usepackage[latin9]{inputenc}
\usepackage{amsmath}
\usepackage{graphicx}
\usepackage{esint}

\makeatletter
\usepackage{amsfonts}
\usepackage{aecompl}\usepackage{epstopdf}

\usepackage{babel}
\date{\today}

\usepackage{babel} 

\makeatother

\usepackage{babel}
\begin{document}

\title{Segregation-induced phase transformations in grain boundaries}

\author{\noindent T. Frolov}

\address{Department of Materials Science and Engineering, University of California,
Berkeley, California 94720, USA}

\author{\noindent M. Asta}

\address{Department of Materials Science and Engineering, University of California,
Berkeley, California 94720, USA}

\author{\noindent Y. Mishin}

\address{Department of Physics and Astronomy, MSN 3F3, George Mason University,
Fairfax, Virginia 22030, USA}
\begin{abstract}
Phase transformations in metallic grain boundaries (GBs) present significant
fundamental interest in the context of thermodynamics of low-dimensional
physical systems. We report on atomistic computer simulations of the
Cu-Ag system that provide direct evidence that GB phase transformations
in a single-component GB can continue to exist in a binary alloy.
This gives rise to segregation-induced phase transformations with
varying chemical composition at a fixed temperature. Furthermore,
for such transformations we propose an approach to calculations of
free energy differences between different GB phases by thermodynamic
integration along a segregation isotherm. This opens the possibility
of developing quantitative thermodynamics of GB phases, their transformations
to each other, and critical phenomena in the future.
\end{abstract}



\maketitle

\section{Introduction}

\emph{Motivation. --} Recent years have seen a rising interest in
phase transformations in two-dimensional systems such as grain boundaries
(GBs) and other interfaces. Experimentally, a number GB phases with
discrete thickness (single layer, bilayer, etc.) have been found in
binary and multi-component metallic systems.\cite{Ma2012,Luo23092011,Cantwell-2013}
Different GB phases have also been observed in ceramic materials,
where they are referred to as intergranular thin films or \textquoteleft \textquoteleft complexions\textquoteright \textquoteright .\cite{Dillon2007,Cantwell-2013,Kaplan2013}
Recently, a methodology has been developed for identification and
structural characterization of GB phases in single-component metallic
systems by atomistic computer simulations.\cite{Frolov2013,Frolov2013a}
Applying this methodology, several different phases and reversible
temperature-induced transformations between them were discovered in
symmetrical tilt $\Sigma5(210)[001]$ and $\Sigma5(310)[001]$ GBs
in face centered cubic metals, $\Sigma$ being the reciprocal density
of coincidence sites, {[}001{]} the tilt axis, and (210) and (310)
the GB planes. For the $\Sigma5(310)[001]$ GB in Cu, Ag GB diffusion
coefficients computed separately for two GB phases\cite{Frolov2013a}
accurately reproduced the break in the temperature dependence of Ag
GB diffusion coefficients measured experimentally by the radio-tracer
method.\cite{Divinski2012} This agreement provided a convincing experimental
evidence for the existence of phase transformations in metallic GBs.
Furthermore, such transformations have been shown to have a strong
effect on shear-coupled GB motion and shear strength.\cite{Frolov:2014aa}

When studying Ag impurity diffusion in the Cu $\Sigma5(310)[001]$
GB,\cite{Frolov2013a} it was found that two GB phases stable at low
and high temperatures displayed different segregation patterns, namely,
a single layer and bilayer, respectively. However, the effect of segregation
on the GB transformation temperature has not been studied either in
Ref.~\onlinecite{Frolov2013a} or in any previous work. Furthermore,
while the proposed methodology\cite{Frolov2013,Frolov2013a} enables
direct observation of GB phase transformations by atomistic simulations,
there have been no experimental or theoretical \emph{quantitative}
estimates of free energy differences between different GB phases. 

The goal of this paper is to report on atomistic simulations of a
segregation-induced GB phase transformation in the Cu-Ag system, and
to demonstrate an approach to calculations of free energy differences
between GB phases. Although we were able to detect segregation-induced
transitions in both $\Sigma5$ GBs mentioned above, in this paper
we focus the attention on the $\Sigma5(210)[001]$ GB, which is different
from the boundary studied earlier.\cite{Frolov2013a} This choice
was dictated by the higher transformation temperature in this boundary,
permitting equilibration of its structure on shorted time scales.
However, the proposed approach can be readily extended to other GBs
in the future. 

\emph{Methodology. --} Our main simulation methods are molecular dynamics
(MD) and semi-grand canonical Monte Carlo (MC) simulations with Cu-Ag
interactions described by an embedded-atom potential.\cite{Williams06}
The MD simulations utilized the LAMMPS code\cite{Plimpton95}, while
for the MC simulations we used the parallel MC algorithm developed
by Sadigh et al.\cite{Sadigh2012} The latter alternates MC switches
of atomic species (swaps) with MD runs implemented in LAMMPS. In this
work, the fraction of swapped atoms was chosen to be 0.3 and the MD
runs between the MC swaps comprised 1,000 integration steps 0.2 fs
each. In the MC simulations, the temperature $T$ and diffusion potential
$M$ of Ag relative to Cu are fixed while the distribution of Ag atoms
over the system can vary to reach thermodynamic equilibrium. Typical
simulation times were between 100 and 280 ns measured by the total
number of MD steps. Images of the GB structures were produced with
the visualization tool AtomEye.\cite{AtomEye}

Two types of simulation block were created. For finding the GB phase
transformation point, the block contained a GB terminated at two free
surfaces normal the {[}120{]} direction $z$. In the $x$ and $y$
directions aligned with the {[}001{]} and {[}210{]} axes, respectively,
the boundary conditions were periodic. (Due to the periodic condition
in $y$, the block effectively contained two GBs.) The system contained
101,031 atoms and had the dimensions of $5.1\times16.5\times150$
nm$^{3}$. The open surfaces serve as sinks and sources of atoms that
can penetrate in and out of the boundary to adjust its local atomic
density to reach equilibrium. It has been previously shown\cite{Frolov2013}
that this simulation setup permits observation of GB phase transformations
in elemental systems. In this work, this simulation approach is extended
to segregation-induced phase transformations in binary systems. 

For computing the amount of GB segregation, two smaller blocks were
created by carving rectangular regions out of the larger system. Each
of the smaller blocks contained a single GB phase and had periodic
boundary conditions in all three directions. The numbers of atoms
in the blocks containing the split kite and filled kite phases were,
respectively, 33,677 and 33,576. Because this construction isolated
the GB phases from sinks and sources of atoms, the GB was unable to
vary its local density and thus maintained its phase during the subsequent
simulations. This enabled us to study thermodynamic properties of
the two phases individually over a range of compositions.

\begin{figure}
\noindent \begin{centering}
\includegraphics[clip,width=0.7\textwidth]{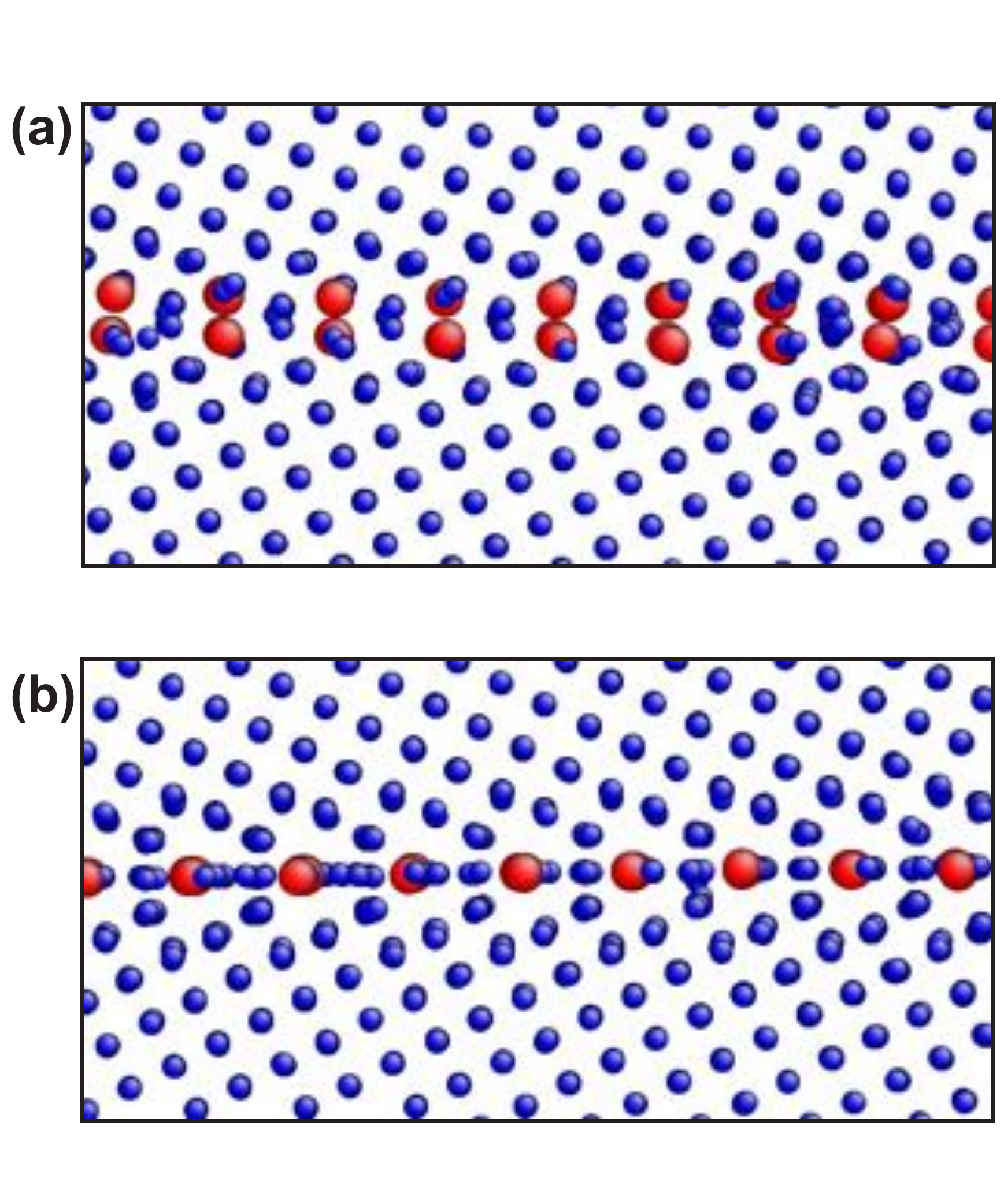}
\par\end{centering}

\protect\caption{Ag GB segregation patterns in the (a) split kite phase and (b) filled
kite phase of the Cu $\Sigma5(210)[001]$. The tilt axis {[}001{]}
is normal to the page. The smaller blue and larger red spheres represent
Cu and Ag atoms, respectively. The segregation formed at the temperature
of $T=900$ K and diffusion potential $M=0.9$ eV. The images are
shown after a short MD run at $T=10$ K to remove thermal noise. \label{fig:Segregation pattern}}
\end{figure}

\emph{Results. --} During the Monte Carlo simulations with open surfaces
at given $T$ and $M$, it was found that after a long run the GB
always reaches equilibrium with a structure of one of the two phases.
The two phases exhibit different segregation patterns: while in the
filled kite phase the Ag atoms segregate to the GB plane as a single
(but not complete) layer, in the split-kite structure they form a
bilayer. The two segregation patterns are illustrated in Fig.~\ref{fig:Segregation pattern}.

\begin{figure}
\noindent \begin{centering}
\includegraphics[clip,width=0.9\textwidth]{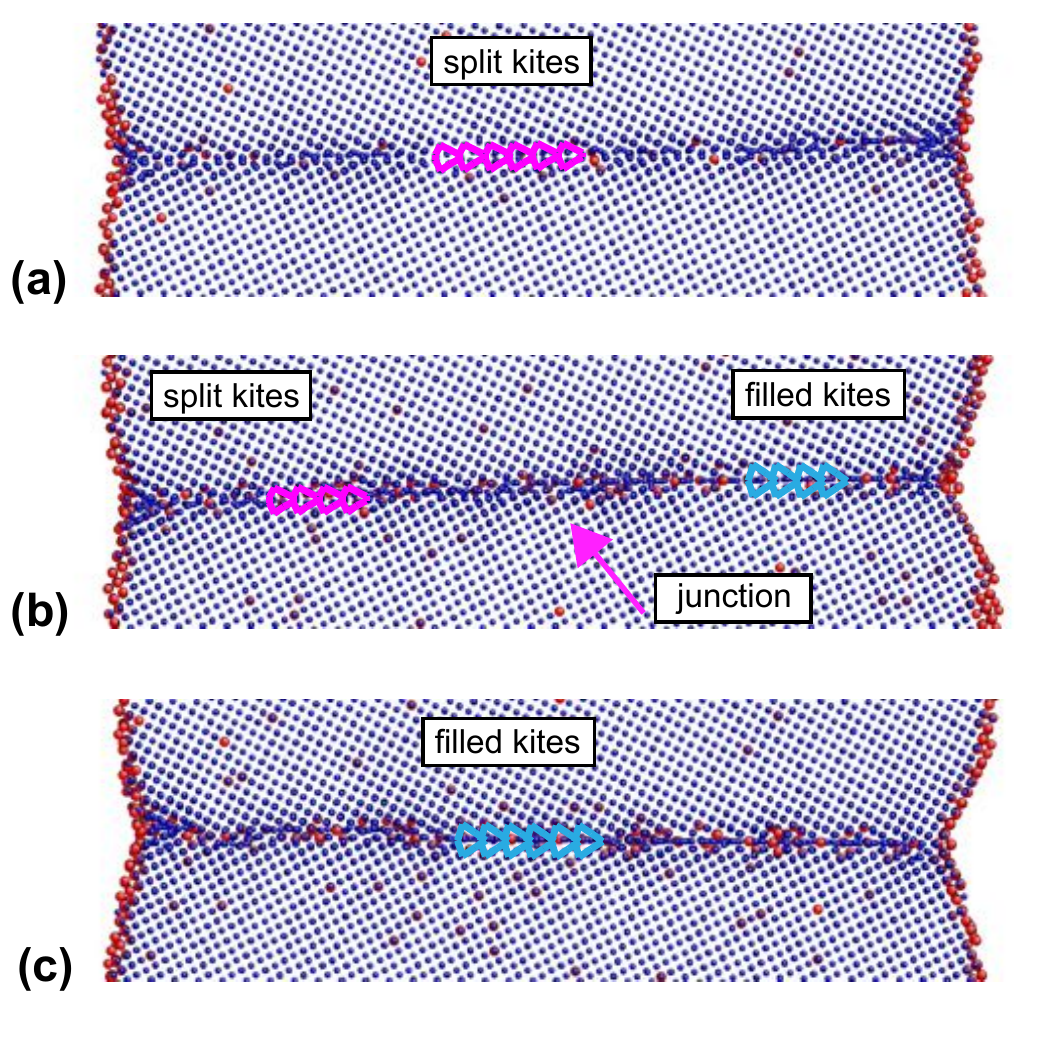}
\par\end{centering}

\protect\caption{GB phase transformation at $T=900$ K and $M=0.48$ eV. (a) Initial
GB phase is split kites (b) Nucleation of the filled kite phase at
the surface. (c) The GB has completely transformed to the filled kite
phase. \label{fig:snaphots of trans }}
\end{figure}

When $M$ changes at a fixed temperature, the GB either re-equilibrates
to a new state of the same phase or transforms to another phase. As
an example, Fig.~\ref{fig:snaphots of trans } illustrates a phase
transformation at the temperature of 900 K. The initial state of the
boundary is split kites created by a previous MD run. The diffusion
potential is then switched to $M=0.48$ eV and a new run is started.
During this run, the filled kite phase nucleates at the GB/surface
junction and begins to grow into the boundary converting its structure
to filled kites. The boundary between the two GB phases is a line
defect that can be considered to be a two-dimensional analog of inter-phase
boundaries in bulk thermodynamics. This two-dimensional inter-phase
boundary penetrates into the GB, and after a 100 ns long simulation
reaches the opposing surface, converting the entire boundary into
the filled kite phase.

To find the equilibrium state between the two GB phases, simulations
were performed for a set of diffusion potentials starting with a GB
containing both phases {[}e.g., Fig.~\ref{fig:snaphots of trans }(b){]}.
It was determined that in all simulations with $M>M*=0.33$ eV, the
inter-phase boundary moved in one direction and the GB ended up in
the split kite phase. In simulations with $M<M*$, the inter-phase
boundary moved in the opposite direction converted the GB to the filled
kite phase. It was, thus, concluded that at this particular temperature,
the two GB phases coexist at $M=M*$. In simulations with exactly
this value of the diffusion potential, both GB structures continued
to coexist during a 280 ns long simulation run, the longest that we
could afford with available computational resources. The obtained
equilibrium value of $M$ corresponds to the grain composition of
$c=0.02$ at.\%Ag. Since the two phases are in equilibrium at this
composition, their GB free energies are equal:
\begin{equation}
\gamma^{SK}=\gamma^{FK}\equiv\gamma_{*}.\label{eq:1}
\end{equation}

GB segregation in individual GB phases was computed in periodic simulation
blocks containing a single phase. As in previous work,\cite{Frolov2012b}
the amount of segregation was defined as the excess number $N_{Ag}$
of Ag atoms per unit GB area relative to a perfect lattice region
with the same composition as the grains and containing the same total
number $N$ of atoms as the bicrystal. This type of segregation is
denoted $[N_{Ag}]_{N}$ and measured in $\textrm{\AA}^{-2}$. It was
computed by averaging over 250 snapshots saved after every ten MC
swaps. 

\begin{figure}
\noindent \begin{centering}
\includegraphics[clip,height=0.8\textheight]{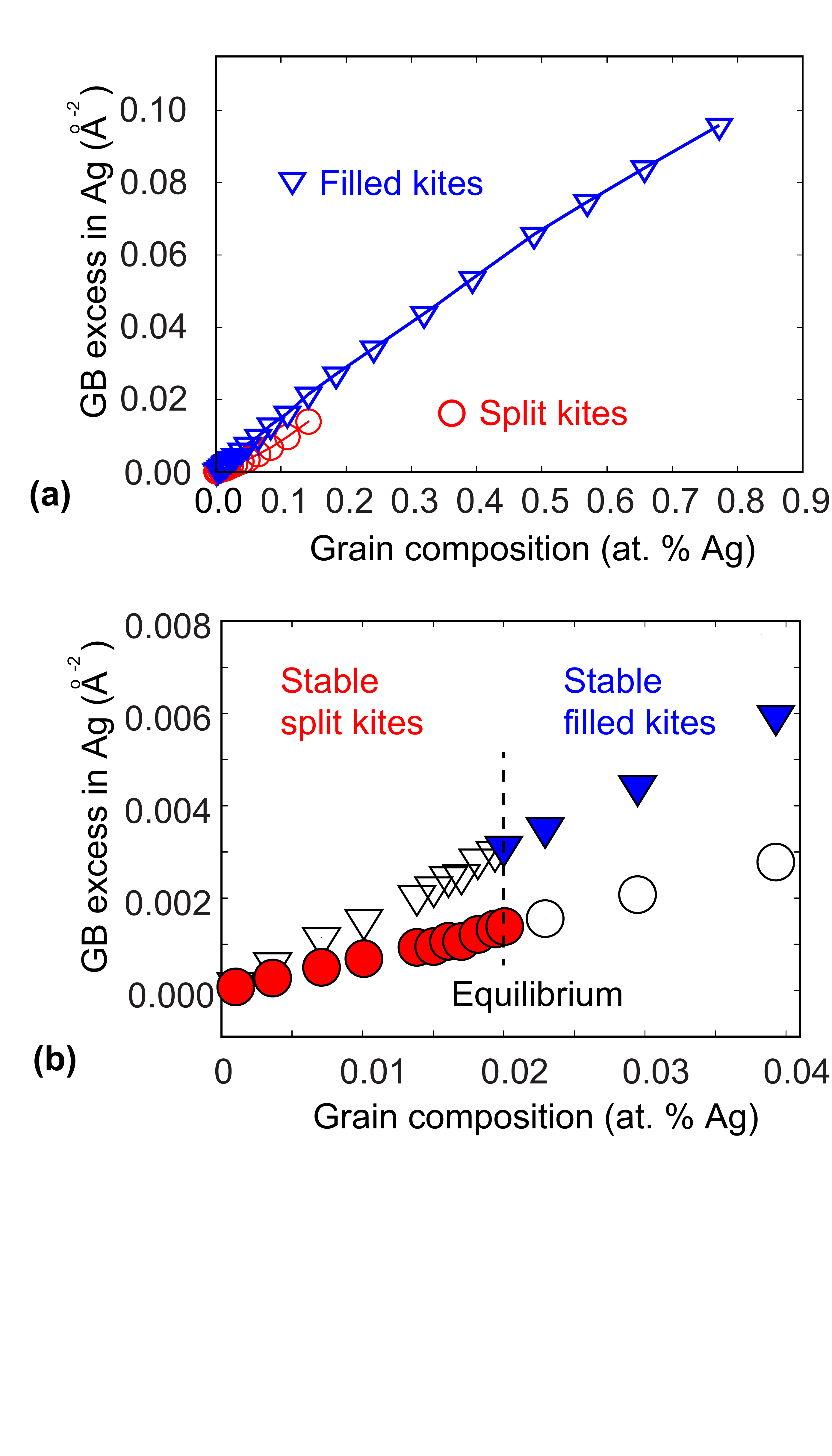}
\par\end{centering}

\protect\caption{Ag GB segregation isotherms for the split kite (circles) and filled
kite (triangle) phases at the temperature of 900 K. The GB excess
of Ag atoms per unit area is plotted as a function of Ag concentration
in the grains. (a) Summary of all results. The curves end when the
GB phase undergoes a disordering transition. (b) Zoomed view of the
composition range near the GB phase transformation at $c=0.02$ at.\%Ag.
The filled and open symbols represent data for the stable and metastable
states, respectively. \label{fig:segregation}}
\end{figure}

Fig.~\ref{fig:segregation} presents the segregation isotherms computed
for individual phases at 900 K. Each curve stops at a point where
the GB structure becomes too disordered to identify it unambiguously
with a particular phase. Note that segregation in the filled kite
phase is systematically higher than in the split kite phase, even
though the former exhibits a bilayer segregation while the latter
a single-layer segregations. Thus, caution should be exercised in
the interpretation of experimental images of segregated GBs: a bilayer
segregation pattern is not necessarily an indication of stronger segregation.
In the zoomed view of this plot displayed in Fig.~\ref{fig:segregation}(b),
the filled circles represent grain compositions for which the GB transformed
to split kites. Likewise, the filled triangles represent grain compositions
for which the GB transformed to filled kites. The composition of $c=0.02$
at.\%Ag marked by the dashed line separates the intervals of thermodynamic
stability of the two phases and is identified with the point of GB
phase coexistence. At this point, the amount of segregation jumps
discontinuously from $[N_{Ag}]_{N}=0.0014$ $\textrm{\AA}^{-2}$ to
$[N_{Ag}]_{N}=0.0031$ $\textrm{\AA}^{-2}$.

The obtained isotherms (Fig.~\ref{fig:segregation}) contain all
information needed for calculations of free energies of the GB phases
at this temperature. Indeed, for each phase, the GB free energy $\gamma$
follows the adsorption equation\cite{Frolov2012b} 
\begin{equation}
d\gamma=-[S]_{N}dT-[N_{Ag}]_{N}dM+\sum_{i,j=1,2}(\tau_{ij}-\gamma)de_{ij},\label{eq:AE}
\end{equation}
where $[S]_{N}$ is the excess entropy defined the same way as the
segregation, $\tau_{ij}$ is the GB stress, and $e_{ij}$ is the lateral
strain tensor parallel to the GB plane. Eq.~(\ref{eq:AE}) is a particular
case of a more general adsorption equation derived in the previous
work.\cite{Frolov2012b} For the present case, this equation has been
simplified due to the absence of applied mechanical stresses. Furthermore,
given the narrow composition interval around the phase transformation,
the lateral strain is extremely small and the last term in Eq.~(\ref{eq:AE})
can be neglected. As a result, Eq.~(\ref{eq:AE}) can be integrated
with respect to $M$ at a fixed temperature to obtain the free energy
of each GB phase relative to their common value $\gamma_{*}$:
\begin{equation}
\gamma-\gamma_{*}=-\intop_{M_{*}}^{M}[N_{Ag}]_{N}dM.\label{eq:3}
\end{equation}
This free energy is a function of $M$ but can be readily converted
to a function of $c$.

\begin{figure}
\noindent \begin{centering}
\includegraphics[clip,width=0.9\textwidth]{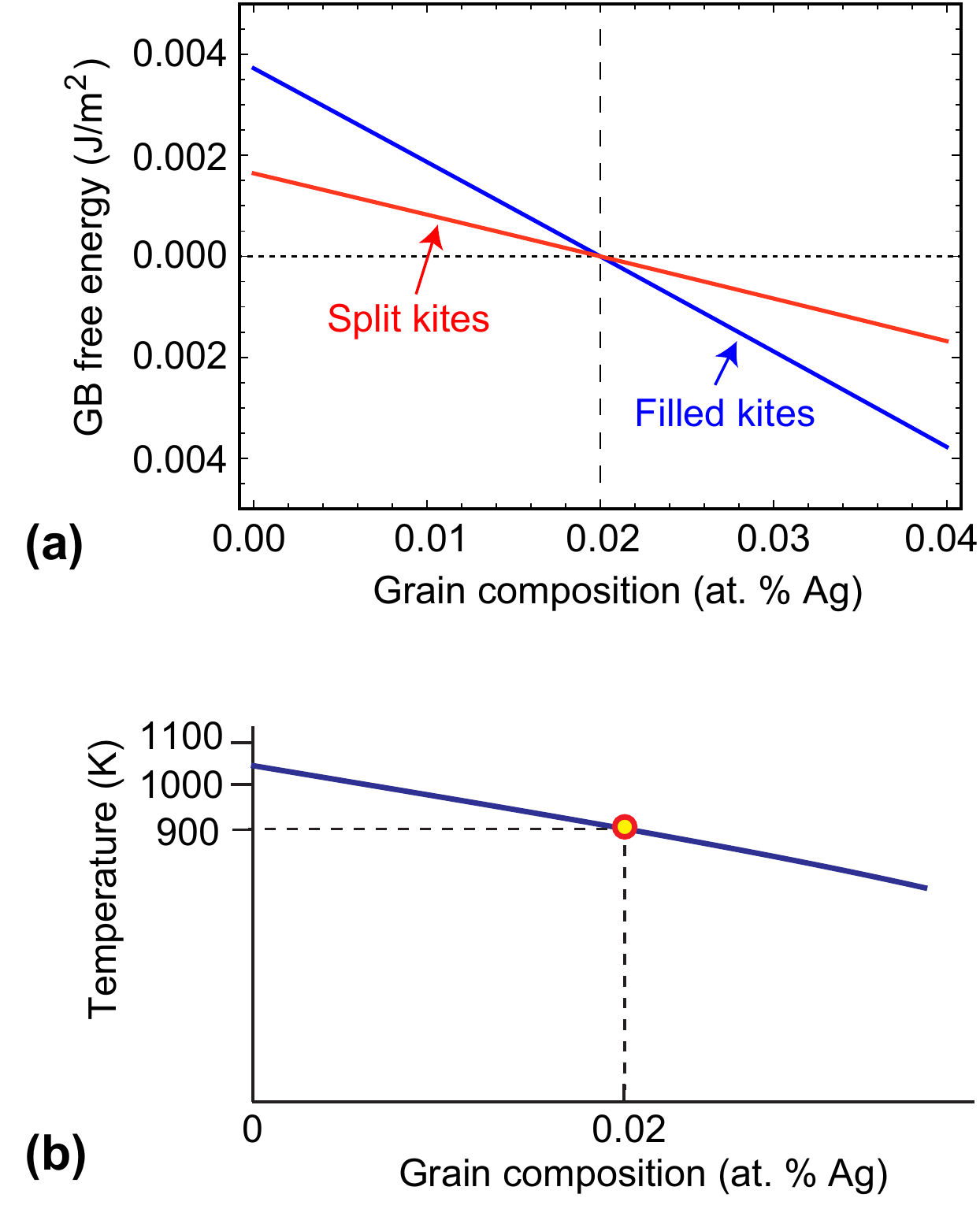}
\par\end{centering}

\protect\caption{(a) Free energies of two GB phases, $\gamma^{SK}$ and $\gamma^{FK}$,
relative to their coexistence value $\gamma_{*}$ as functions of
Ag concentration in the grains. The phase transformation occurs at
$c=0.02$ at.\%Ag. The results were obtained by thermodynamic integration
at 900 K. (b) Schematic phase diagram of GB phase transformations
with the open circle showing the result of this work. \label{fig:gamma}}
\end{figure}

The function $M(c)$ is known from MC simulations of the perfect lattice.
For perform the integration, this function was fitted by the expression
$M(c)=a_{0}+a_{1}c+a_{2}c^{2}+a_{3}log(c)$ with adjustable coefficients
$a_{i}$. Likewise, the segregation isotherms (Fig.~\ref{fig:segregation})
were fitted by $[N_{Ag}]_{N}(c)=b_{1}c+b_{2}c^{2}$ with adjustable
coefficients $b_{i}$ for each phase. After this, the integration
was executed analytically on either side of the equilibrium point,
including extrapolation to pure Cu ($c=0$). The obtained GB free
energies, $\gamma^{SK}-\gamma_{*}$ and $\gamma^{FK}-\gamma_{*}$,
are plotted in Fig.~\ref{fig:gamma}(a) as functions of grain composition.
The striking observation is that the free energy difference between
the two phases is very small. Even in the pure Cu limit, this difference
is as small as 2.2 mJ/m$^{2}$ (compare with the 0 K energy of this
boundary, 0.951 J/m$^{2}$).\cite{Frolov2013} Near the phase equilibrium
point ($c=0.02$ at.\%Ag), the phase transformation can still be reliably
detected when $\gamma^{SK}-\gamma^{FK}$ is less than 0.1 mJ/m$^{2}$!
These numbers demonstrate that using the proposed simulation methods,
thermodynamic properties of GB phases can be characterized with a
high precision, including accurate location of phase transformation
points.

For pure Cu, the energy difference between the two GB phases at 0
K is $\gamma^{FK}-\gamma^{SK}=17$ mJ/m$^{2}$.\cite{Frolov2013}
The respective free energy difference at 900 K is 2.2 mJ/m$^{2}$,
i.e., a factor of seven less. This is consistent with the proximity
of temperature-induced phase transformation in this GB, which was
previously found to occur around 1050 K.\cite{Frolov2013} 

\emph{Conclusions. --} We have demonstrated that the temperature-induced
GB phase transformations found in previous work\cite{Frolov2013}
continue to exist in the presence of segregating solute atoms, giving
rise to segregation-induced GB phase transformations. Such transformations
are accompanied by a discontinuous jump in the amount of segregation
and a change in the segregation pattern from a single layer to a bilayer.
We have studied this transformation at a single temperature of 900
K. However, considering the small solute concentration causing this
transformation, it is likely to be a point on a phase transformation
line that can be conveniently shown in coordinates temperature-composition
{[}Fig.~\ref{fig:gamma}(b){]}. This line terminates at the temperature
axis at around $T=1050$ K. Another segregation-induced structural
transformation was previously found in a twist GB in the Ni-Pt system.\cite{Seidman96}
However, by contrast to the present work, there was no evidence that
a similar transformation occurs in the pure Ni boundary. Most likely,
that transformation represents an isolated region on the temperature-composition
phase diagram that is not connected to the temperature axis. These
two cases suggests that future atomistic simulations may reveal a
rich variety of congruent\cite{Cahn82a} GB phase transformations,
which can be presented as $T$-$c$ phase diagrams by analogy with
bulk systems. 

The thermodynamic integration scheme proposed in this work enables
free energy calculations for individual GB phases (relative to their
common value at equilibrium) and construction of GB phase diagrams.
In particular, the entire phase transformation line shown schematically
in Fig.~\ref{fig:segregation}(b) can be calculated by the same method.
This line could be tested against predictions of interface thermodynamics.
Because Eq.~(\ref{eq:1}) must remain valid along this line, we can
use the adsorption equation (\ref{eq:AE}) for each phase to derive
the following equation for the slope of the equilibrium line:
\begin{equation}
\dfrac{dT}{dc}=-\dfrac{\Delta[N_{Ag}]_{N}\left(\partial M/\partial c\right)_{T}T}{\Delta[U]_{N}-M\Delta[N_{Ag}]_{N}+\Delta[N_{Ag}]_{N}\left(\partial M/\partial T\right)_{c}T},\label{eq:slope}
\end{equation}
where $[U]_{N}$ is the excess internal energy of the GB and symbol
$\Delta$ denotes differences between properties of the two phases.
The quantities appearing in the right-hand side can be computed separately
to compare the right-hand side with the actual slope of the line.

The proposed method can be applied for calculations of not only phase
equilibrium lines but also spinodal lines and critical points. For
example, the obtained segregation isotherms {[}Fig.~\ref{fig:segregation}(a){]}
terminate at points of GB disorder. While investigation of the nature
of this disordering transition is beyond the scope of this paper,
it is possible that these points signify transformations to new phases
that are yet to be discovered. Investigation of critical phenomena
in two-dimensional GB phases may present significant fundamental interest. 

\vspace{0.15in}

T.F. was supported by a post-doctoral fellowship from the Miller Institute
for Basic Research in Science at University of California, Berkeley.
Y.M. was supported by the National Science Foundation, Division of
Materials Research, the Metals and Metallic Nanostructures Program.

%\bibliography{/Users/ymishin/YURI/Bibliography/literat}

\begin{thebibliography}{16}%
\makeatletter
\providecommand \@ifxundefined [1]{%
 \@ifx{#1\undefined}
}%
\providecommand \@ifnum [1]{%
 \ifnum #1\expandafter \@firstoftwo
 \else \expandafter \@secondoftwo
 \fi
}%
\providecommand \@ifx [1]{%
 \ifx #1\expandafter \@firstoftwo
 \else \expandafter \@secondoftwo
 \fi
}%
\providecommand \natexlab [1]{#1}%
\providecommand \enquote  [1]{``#1''}%
\providecommand \bibnamefont  [1]{#1}%
\providecommand \bibfnamefont [1]{#1}%
\providecommand \citenamefont [1]{#1}%
\providecommand \href@noop [0]{\@secondoftwo}%
\providecommand \href [0]{\begingroup \@sanitize@url \@href}%
\providecommand \@href[1]{\@@startlink{#1}\@@href}%
\providecommand \@@href[1]{\endgroup#1\@@endlink}%
\providecommand \@sanitize@url [0]{\catcode `\\12\catcode `\$12\catcode
  `\&12\catcode `\#12\catcode `\^12\catcode `\_12\catcode `\%12\relax}%
\providecommand \@@startlink[1]{}%
\providecommand \@@endlink[0]{}%
\providecommand \url  [0]{\begingroup\@sanitize@url \@url }%
\providecommand \@url [1]{\endgroup\@href {#1}{\urlprefix }}%
\providecommand \urlprefix  [0]{URL }%
\providecommand \Eprint [0]{\href }%
\providecommand \doibase [0]{http://dx.doi.org/}%
\providecommand \selectlanguage [0]{\@gobble}%
\providecommand \bibinfo  [0]{\@secondoftwo}%
\providecommand \bibfield  [0]{\@secondoftwo}%
\providecommand \translation [1]{[#1]}%
\providecommand \BibitemOpen [0]{}%
\providecommand \bibitemStop [0]{}%
\providecommand \bibitemNoStop [0]{.\EOS\space}%
\providecommand \EOS [0]{\spacefactor3000\relax}%
\providecommand \BibitemShut  [1]{\csname bibitem#1\endcsname}%
\let\auto@bib@innerbib\@empty
%</preamble>
\bibitem [{\citenamefont {Ma}\ \emph {et~al.}(2012)\citenamefont {Ma},
  \citenamefont {Asl}, \citenamefont {Tansarawiput}, \citenamefont {Cantwell},
  \citenamefont {Qi}, \citenamefont {Harmer},\ and\ \citenamefont
  {Luo}}]{Ma2012}%
  \BibitemOpen
  \bibfield  {author} {\bibinfo {author} {\bibfnamefont {S.}~\bibnamefont
  {Ma}}, \bibinfo {author} {\bibfnamefont {K.~M.}\ \bibnamefont {Asl}},
  \bibinfo {author} {\bibfnamefont {C.}~\bibnamefont {Tansarawiput}}, \bibinfo
  {author} {\bibfnamefont {P.~R.}\ \bibnamefont {Cantwell}}, \bibinfo {author}
  {\bibfnamefont {M.}~\bibnamefont {Qi}}, \bibinfo {author} {\bibfnamefont
  {M.~P.}\ \bibnamefont {Harmer}}, \ and\ \bibinfo {author} {\bibfnamefont
  {J.}~\bibnamefont {Luo}},\ }\bibfield  {title} {\enquote {\bibinfo {title} {A
  grain boundary phase transition in {Si--Au}},}\ }\href@noop {} {\bibfield
  {journal} {\bibinfo  {journal} {Scripta Mater.}\ }\textbf {\bibinfo {volume}
  {66}},\ \bibinfo {pages} {203--206} (\bibinfo {year} {2012})}\BibitemShut
  {NoStop}%
\bibitem [{\citenamefont {Luo}\ \emph {et~al.}(2011)\citenamefont {Luo},
  \citenamefont {Cheng}, \citenamefont {Asl}, \citenamefont {Kiely},\ and\
  \citenamefont {Harmer}}]{Luo23092011}%
  \BibitemOpen
  \bibfield  {author} {\bibinfo {author} {\bibfnamefont {Jian}\ \bibnamefont
  {Luo}}, \bibinfo {author} {\bibfnamefont {Huikai}\ \bibnamefont {Cheng}},
  \bibinfo {author} {\bibfnamefont {Kaveh~Meshinchi}\ \bibnamefont {Asl}},
  \bibinfo {author} {\bibfnamefont {Christopher~J.}\ \bibnamefont {Kiely}}, \
  and\ \bibinfo {author} {\bibfnamefont {Martin~P.}\ \bibnamefont {Harmer}},\
  }\bibfield  {title} {\enquote {\bibinfo {title} {The role of a bilayer
  interfacial phase on liquid metal embrittlement},}\ }\href {\doibase
  10.1126/science.1208774} {\bibfield  {journal} {\bibinfo  {journal}
  {Science}\ }\textbf {\bibinfo {volume} {333}},\ \bibinfo {pages} {1730--1733}
  (\bibinfo {year} {2011})}\BibitemShut {NoStop}%
\bibitem [{\citenamefont {Cantwell}\ \emph {et~al.}(2013)\citenamefont
  {Cantwell}, \citenamefont {Tang}, \citenamefont {Dillon}, \citenamefont
  {Luo}, \citenamefont {Rohrer},\ and\ \citenamefont {Harmer}}]{Cantwell-2013}%
  \BibitemOpen
  \bibfield  {author} {\bibinfo {author} {\bibfnamefont {P.~R.}\ \bibnamefont
  {Cantwell}}, \bibinfo {author} {\bibfnamefont {M.}~\bibnamefont {Tang}},
  \bibinfo {author} {\bibfnamefont {S.~J.}\ \bibnamefont {Dillon}}, \bibinfo
  {author} {\bibfnamefont {J.}~\bibnamefont {Luo}}, \bibinfo {author}
  {\bibfnamefont {G.~S.}\ \bibnamefont {Rohrer}}, \ and\ \bibinfo {author}
  {\bibfnamefont {M.~P.}\ \bibnamefont {Harmer}},\ }\bibfield  {title}
  {\enquote {\bibinfo {title} {Grain boundary complexions},}\ }\href@noop {}
  {\bibfield  {journal} {\bibinfo  {journal} {Acta Mater.}\ }\textbf {\bibinfo
  {volume} {62}},\ \bibinfo {pages} {1--48} (\bibinfo {year}
  {2013})}\BibitemShut {NoStop}%
\bibitem [{\citenamefont {Dillon}\ \emph {et~al.}(2007)\citenamefont {Dillon},
  \citenamefont {Tang}, \citenamefont {Carter},\ and\ \citenamefont
  {Harmer}}]{Dillon2007}%
  \BibitemOpen
  \bibfield  {author} {\bibinfo {author} {\bibfnamefont {S.~J.}\ \bibnamefont
  {Dillon}}, \bibinfo {author} {\bibfnamefont {M.}~\bibnamefont {Tang}},
  \bibinfo {author} {\bibfnamefont {W.~C.}\ \bibnamefont {Carter}}, \ and\
  \bibinfo {author} {\bibfnamefont {M.~P.}\ \bibnamefont {Harmer}},\ }\bibfield
   {title} {\enquote {\bibinfo {title} {Complexion: A new concept for kinetic
  engineering in materials science},}\ }\href@noop {} {\bibfield  {journal}
  {\bibinfo  {journal} {Acta Mater.}\ }\textbf {\bibinfo {volume} {55}},\
  \bibinfo {pages} {6208--6218} (\bibinfo {year} {2007})}\BibitemShut {NoStop}%
\bibitem [{\citenamefont {Kaplan}\ \emph {et~al.}(2013)\citenamefont {Kaplan},
  \citenamefont {Chatain}, \citenamefont {Wynblatt},\ and\ \citenamefont
  {Carter}}]{Kaplan2013}%
  \BibitemOpen
  \bibfield  {author} {\bibinfo {author} {\bibfnamefont {Wayne~D.}\
  \bibnamefont {Kaplan}}, \bibinfo {author} {\bibfnamefont {Dominique}\
  \bibnamefont {Chatain}}, \bibinfo {author} {\bibfnamefont {Paul}\
  \bibnamefont {Wynblatt}}, \ and\ \bibinfo {author} {\bibfnamefont {W.~Craig}\
  \bibnamefont {Carter}},\ }\bibfield  {title} {\enquote {\bibinfo {title} {A
  review of wetting versus adsorption, complexions, and related phenomena: the
  rosetta stone of wetting},}\ }\href@noop {} {\bibfield  {journal} {\bibinfo
  {journal} {J. Mater. Sci.}\ }\textbf {\bibinfo {volume} {48}},\ \bibinfo
  {pages} {5681--5717} (\bibinfo {year} {2013})}\BibitemShut {NoStop}%
\bibitem [{\citenamefont {Frolov}\ \emph
  {et~al.}(2013{\natexlab{a}})\citenamefont {Frolov}, \citenamefont {Olmsted},
  \citenamefont {Asta},\ and\ \citenamefont {Mishin}}]{Frolov2013}%
  \BibitemOpen
  \bibfield  {author} {\bibinfo {author} {\bibfnamefont {T.}~\bibnamefont
  {Frolov}}, \bibinfo {author} {\bibfnamefont {D.~L.}\ \bibnamefont {Olmsted}},
  \bibinfo {author} {\bibfnamefont {M.}~\bibnamefont {Asta}}, \ and\ \bibinfo
  {author} {\bibfnamefont {Y.}~\bibnamefont {Mishin}},\ }\bibfield  {title}
  {\enquote {\bibinfo {title} {Structural phase transformations in metallic
  grain boundaries},}\ }\href@noop {} {\bibfield  {journal} {\bibinfo
  {journal} {Nature Communications}\ }\textbf {\bibinfo {volume} {4}},\
  \bibinfo {pages} {1899} (\bibinfo {year} {2013}{\natexlab{a}})}\BibitemShut
  {NoStop}%
\bibitem [{\citenamefont {Frolov}\ \emph
  {et~al.}(2013{\natexlab{b}})\citenamefont {Frolov}, \citenamefont {Divinski},
  \citenamefont {Asta},\ and\ \citenamefont {Mishin}}]{Frolov2013a}%
  \BibitemOpen
  \bibfield  {author} {\bibinfo {author} {\bibfnamefont {T.}~\bibnamefont
  {Frolov}}, \bibinfo {author} {\bibfnamefont {S.~V.}\ \bibnamefont
  {Divinski}}, \bibinfo {author} {\bibfnamefont {M.}~\bibnamefont {Asta}}, \
  and\ \bibinfo {author} {\bibfnamefont {Y.}~\bibnamefont {Mishin}},\
  }\bibfield  {title} {\enquote {\bibinfo {title} {Effect of interface phase
  transformations on diffusion and segregation in high-angle grain
  boundaries},}\ }\href@noop {} {\bibfield  {journal} {\bibinfo  {journal}
  {Phys. Rev. Lett.}\ }\textbf {\bibinfo {volume} {110}},\ \bibinfo {pages}
  {255502} (\bibinfo {year} {2013}{\natexlab{b}})}\BibitemShut {NoStop}%
\bibitem [{\citenamefont {Divinski}\ \emph {et~al.}(2012)\citenamefont
  {Divinski}, \citenamefont {Edelhoff},\ and\ \citenamefont
  {Prokofjev}}]{Divinski2012}%
  \BibitemOpen
  \bibfield  {author} {\bibinfo {author} {\bibfnamefont {S.~V.}\ \bibnamefont
  {Divinski}}, \bibinfo {author} {\bibfnamefont {H.}~\bibnamefont {Edelhoff}},
  \ and\ \bibinfo {author} {\bibfnamefont {S.}~\bibnamefont {Prokofjev}},\
  }\bibfield  {title} {\enquote {\bibinfo {title} {Diffusion and segregation of
  silver in copper {$\Sigma 5$ (310)} grain boundary},}\ }\href@noop {}
  {\bibfield  {journal} {\bibinfo  {journal} {Phys. Rev. {\rm B}}\ }\textbf
  {\bibinfo {volume} {85}},\ \bibinfo {pages} {144104} (\bibinfo {year}
  {2012})}\BibitemShut {NoStop}%
\bibitem [{\citenamefont {Frolov}(2014)}]{Frolov:2014aa}%
  \BibitemOpen
  \bibfield  {author} {\bibinfo {author} {\bibfnamefont {T.}~\bibnamefont
  {Frolov}},\ }\bibfield  {title} {\enquote {\bibinfo {title} {Effect of
  interfacial structural phase transformations on the coupled motion of grain
  boundaries: A molecular dynamics study},}\ }\href@noop {} {\bibfield
  {journal} {\bibinfo  {journal} {Appl. Phys. Lett.}\ }\textbf {\bibinfo
  {volume} {104}},\ \bibinfo {pages} {211905} (\bibinfo {year}
  {2014})}\BibitemShut {NoStop}%
\bibitem [{\citenamefont {Williams}\ \emph {et~al.}(2006)\citenamefont
  {Williams}, \citenamefont {Mishin},\ and\ \citenamefont
  {Hamilton}}]{Williams06}%
  \BibitemOpen
  \bibfield  {author} {\bibinfo {author} {\bibfnamefont {P.~L.}\ \bibnamefont
  {Williams}}, \bibinfo {author} {\bibfnamefont {Y.}~\bibnamefont {Mishin}}, \
  and\ \bibinfo {author} {\bibfnamefont {J.~C.}\ \bibnamefont {Hamilton}},\
  }\bibfield  {title} {\enquote {\bibinfo {title} {An embedded-atom potential
  for the {Cu-Ag} system},}\ }\href@noop {} {\bibfield  {journal} {\bibinfo
  {journal} {Modelling Simul. Mater. Sci. Eng.}\ }\textbf {\bibinfo {volume}
  {14}},\ \bibinfo {pages} {817--833} (\bibinfo {year} {2006})}\BibitemShut
  {NoStop}%
\bibitem [{\citenamefont {Plimpton}(1995)}]{Plimpton95}%
  \BibitemOpen
  \bibfield  {author} {\bibinfo {author} {\bibfnamefont {S.}~\bibnamefont
  {Plimpton}},\ }\bibfield  {title} {\enquote {\bibinfo {title} {Fast parallel
  algorithms for short-range molecular-dynamics},}\ }\href@noop {} {\bibfield
  {journal} {\bibinfo  {journal} {J. Comput. Phys.}\ }\textbf {\bibinfo
  {volume} {117}},\ \bibinfo {pages} {1--19} (\bibinfo {year}
  {1995})}\BibitemShut {NoStop}%
\bibitem [{\citenamefont {Sadigh}\ \emph {et~al.}(2012)\citenamefont {Sadigh},
  \citenamefont {Erhart}, \citenamefont {Stukowski}, \citenamefont {Caro},
  \citenamefont {Martinez},\ and\ \citenamefont
  {\mbox{Zepeda-Ruiz}}}]{Sadigh2012}%
  \BibitemOpen
  \bibfield  {author} {\bibinfo {author} {\bibfnamefont {Babak}\ \bibnamefont
  {Sadigh}}, \bibinfo {author} {\bibfnamefont {Paul}\ \bibnamefont {Erhart}},
  \bibinfo {author} {\bibfnamefont {Alexander}\ \bibnamefont {Stukowski}},
  \bibinfo {author} {\bibfnamefont {Alfredo}\ \bibnamefont {Caro}}, \bibinfo
  {author} {\bibfnamefont {Enrique}\ \bibnamefont {Martinez}}, \ and\ \bibinfo
  {author} {\bibfnamefont {Luis}\ \bibnamefont {\mbox{Zepeda-Ruiz}}},\
  }\bibfield  {title} {\enquote {\bibinfo {title} {Scalable parallel monte
  carlo algorithm for atomistic simulations of precipitation in alloys},}\
  }\href@noop {} {\bibfield  {journal} {\bibinfo  {journal} {Phys. Rev. {\rm
  B}}\ }\textbf {\bibinfo {volume} {85}},\ \bibinfo {pages} {184203} (\bibinfo
  {year} {2012})}\BibitemShut {NoStop}%
\bibitem [{\citenamefont {Li}(2003)}]{AtomEye}%
  \BibitemOpen
  \bibfield  {author} {\bibinfo {author} {\bibfnamefont {J.}~\bibnamefont
  {Li}},\ }\bibfield  {title} {\enquote {\bibinfo {title} {Atomeye: an
  efficient atomistic configuration viewer},}\ }\href@noop {} {\bibfield
  {journal} {\bibinfo  {journal} {Modelling Simul. Mater. Sci. Eng.}\ }\textbf
  {\bibinfo {volume} {11}},\ \bibinfo {pages} {173} (\bibinfo {year}
  {2003})}\BibitemShut {NoStop}%
\bibitem [{\citenamefont {Frolov}\ and\ \citenamefont
  {Mishin}(2012)}]{Frolov2012b}%
  \BibitemOpen
  \bibfield  {author} {\bibinfo {author} {\bibfnamefont {T.}~\bibnamefont
  {Frolov}}\ and\ \bibinfo {author} {\bibfnamefont {Y.}~\bibnamefont
  {Mishin}},\ }\bibfield  {title} {\enquote {\bibinfo {title} {Thermodynamics
  of coherent interfaces under mechanical stresses. {II}. application to
  atomistic simulation of grain boundaries},}\ }\href@noop {} {\bibfield
  {journal} {\bibinfo  {journal} {Phys. Rev. B}\ }\textbf {\bibinfo {volume}
  {85}},\ \bibinfo {pages} {224107} (\bibinfo {year} {2012})}\BibitemShut
  {NoStop}%
\bibitem [{\citenamefont {Udler}\ and\ \citenamefont
  {Seidman}(1996)}]{Seidman96}%
  \BibitemOpen
  \bibfield  {author} {\bibinfo {author} {\bibfnamefont {D.}~\bibnamefont
  {Udler}}\ and\ \bibinfo {author} {\bibfnamefont {D.~N.}\ \bibnamefont
  {Seidman}},\ }\bibfield  {title} {\enquote {\bibinfo {title} {Congruent phase
  transition at a twist boundary induced by solute segregation},}\ }\href@noop
  {} {\bibfield  {journal} {\bibinfo  {journal} {Phys. Rev. Letters}\ }\textbf
  {\bibinfo {volume} {77}},\ \bibinfo {pages} {3379--3382} (\bibinfo {year}
  {1996})}\BibitemShut {NoStop}%
\bibitem [{\citenamefont {Cahn}(1982)}]{Cahn82a}%
  \BibitemOpen
  \bibfield  {author} {\bibinfo {author} {\bibfnamefont {J.~W.}\ \bibnamefont
  {Cahn}},\ }\bibfield  {title} {\enquote {\bibinfo {title} {Transitions and
  phase equilibria among grain boundary structures},}\ }\href@noop {}
  {\bibfield  {journal} {\bibinfo  {journal} {J. Physique Colloques}\ }\textbf
  {\bibinfo {volume} {43}},\ \bibinfo {pages} {199--213} (\bibinfo {year}
  {1982})}\BibitemShut {NoStop}%
\end{thebibliography}

\newpage{}\clearpage{}

\end{document}